\documentclass[onecolumn,showpacs,aps,pre]{revtex4}
\usepackage{graphicx}
\usepackage{bm}
\newcommand{\be}{\begin{equation}}
\newcommand{\ee}{\end{equation}}
\newcommand{\bea}{\begin{eqnarray}}
\newcommand{\eea}{\end{eqnarray}}
\usepackage{dcolumn}
\usepackage{amsmath}
\usepackage{latexsym}

\begin {document}

\title {Large compact clusters and fast dynamics in coupled nonequilibrium systems}
\author{Shauri Chakraborty$^1$, Sukla Pal$^1$, Sakuntala Chatterjee$^1$ and 
Mustansir Barma$^2$}
\affiliation{ $^1$ Department of Theoretical Sciences, S.N. Bose National Centre
for Basic Sciences, Block - JD, Sector - III, Salt Lake, 
Kolkata 700098, India \\ $^2$ TIFR Centre for Interdisciplinary Sciences, 21
Brundavan Colony, Osman Sagar Road, Narsingi, Hyderabad 500075, India.}
\begin{abstract}
We demonstrate particle clustering on macroscopic scales in a coupled nonequilibrium system where two species of particles are advected by a fluctuating landscape and modify the landscape in the process. The phase diagram generated by varying the particle-landscape coupling, valid for all particle density and in both one and two dimensions, shows novel nonequilibrium phases. While particle species are completely phase separated, the landscape develops macroscopically ordered regions coexisting with a disordered region, resulting in coarsening and steady state dynamics on time scales which grow algebraically with size, not seen earlier in systems with pure domains.
\end{abstract}
\maketitle

Particle clustering is important in many natural physical and biological
phenomena, for instance, the formation of sediments \cite{sriram} and protein
clustering on a biological membrane \cite{madan12}. Evidently, it is important to
understand processes that cause clustering in different physical contexts, and
how these processes influence the properties of the cluster and the time taken
to form it.  Often, large-scale clustering is driven by interactions with an
external medium which itself has correlations in space and time
\cite{deutsch,mehlig,drossel}. An
important physical effect in such systems is the back-influence of the particles
on the medium. 
This interaction can aid clustering, or destroy it.
If a cluster does form, it may be compact and robust, or a dynamic object that
undergoes constant reorganization. The formation time may grow exponentially
with the size, or as a power law. Given this wealth of possibilities, it is
important to look for an understanding, within simple models, of the
circumstances under which different sorts of macroscopically clustered states
occur.

In this letter, we derive the phase diagram of a simple model system as we vary
the interaction between the environment and particles. In the process, we unmask
a novel non-equilibrium phase of particles with compact clustering and rich and
rapid dynamics coexisting with a macroscopically organized landscape. The model 
has partial overlap with the lattice gas model
of Lahiri and Ramaswamy (LR) for sedimenting colloidal crystals 
\cite{lahiri97,lahiri}, but the new phases manifest themselves outside the LR 
regime. Our results hold in both one and two dimensions.

\begin{figure}
\includegraphics[scale=0.5,angle=0]{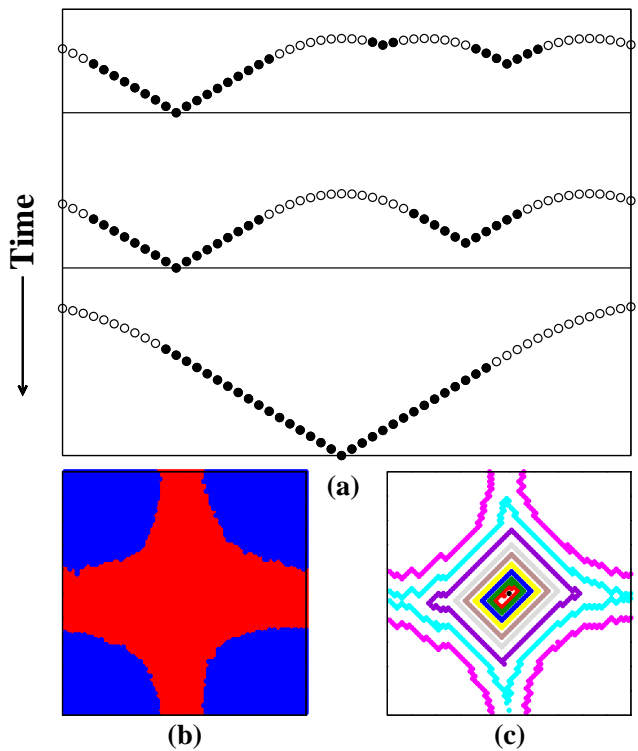}
\caption{Schematic presentation of different phases in one and two dimensions.
(a) shows the coarsening mechanism in one dimension for $b'\leq 0$. 
$H$ ($L$) particles are shown by solid (hollow) circles. (b) and (c) show
typical configurations in two dimensions. In (b) $H$ ($L$) cluster shown by red
(blue) color and (c) shows the equal height contours.}
\label{fig:scheme}
\end{figure}

The model consists of two sets of particles moving stochastically 
in a fluctuating potential
energy landscape. Particles try to minimize their energy
by (a) moving along the local potential gradient of the landscape and (b)
modifying the landscape around them in such a way as to lower the energy
further. The model is generic but we discuss it in the language of 
particles confined to move on a fluctuating surface in the presence of gravity, 
where the particles can locally distort the surface shape
to further lower the energy (see Fig. \ref{fig:scheme}).  
One of the particle species is considered lighter and the other is
heavier; we use the name LH (Light-heavy) model to describe the system. 
Process (b) affects the landscape dynamics
quite differently in parts which are rich or poor in one species of particle,
ultimately resulting in the formation of
distinct macroscopic regions, each corresponding to
a phase. Our study reveals a rich set of phenomena: strong phase separation with
fluctuationless phases for particles, but a different sort of organization
 for the landscape; a rapid
approach to the steady state; and intricate steady state dynamics of the
interfaces between phases, with three distinct temporal regimes.  

There has been a recent surge of activity in the field of coupled driven
diffusive systems \cite{beijeren,ferrari,gunter_pnas}, and it is useful to view
our work in this context. This activity has resulted in a catalogue of
universality classes which describe how propagating modes in these systems decay
in time.  The modes themselves are defined by diagonalizing coupled hydrodynamic
equations to linear order, with eigenvalues giving their speeds of propagation. The
disordered phase of our system is indeed described by this theory. But the
ordered phases of primary interest to us correspond to {\it complex} eigenvalues at
the linear level \cite{cplx}; the
imaginary parts signal instabilities, heralding macroscopic phase separation.
However, such linear analysis cannot reveal the characteristics of the final
phases, which can and do differ from each other in fundamental respects. Our
results provide the necessary characterization and thus contribute to the
important goal of classifying ordered phases in coupled driven diffusive systems.

In a nutshell, the quintessentially nonequilibrium states found here exhibit
phase separation, with qualitatively different types of ordering for the
particles and landscape, quite unlike systems known earlier. In particular,
particles display strong phase separation \cite{lahiri} characterized by
pure, fluctuationless phases, which cohabit with  three macroscopic regions of
the surface, two of which  possess long range order, while the third does not. These
findings differ markedly from the strongly phase separated states found earlier
in the LR and ABC models \cite{lahiri,abc} and imply strong changes for both 
static and dynamical properties. Notably, the approach to our steady state is 
rapid, with a coarsening time that grows as a power law of size, as opposed to 
the much slower exponentially growing time scales found earlier.

The LH model consists of two coupled driven diffusive systems, with
conserved quantities. This is a lattice model of $H$ (heavier) and
$L$ (lighter) particles with damped motion under gravity and residing on a
fluctuating surface. The local dynamics of the particles and the surface are
coupled: $H$ and $L$ particles at neighboring sites may interchange locations, and
do so preferentially if the local tilt of the surface favors a downward move
for $H$. Particles reside on lattice sites and interact via hard-core
exclusion: a site holds at most one particle ($H$ or $L$). 
If the symbols $/ $ and  $ \backslash $ indicate upward and downward tilts of the 
surface,
respectively, then the particles follow the dynamics: 
\begin{eqnarray}
W(H\backslash L \rightarrow L\backslash H)&=& D+a\nonumber\\
W(L \backslash H \rightarrow H\backslash L)&=& D-a\nonumber\\
W(H / L \rightarrow L / H)&=& D-a\nonumber\\
W( L / H \rightarrow H / L)&=& D+a  
\label{eq:model1}
\end{eqnarray}
where $W$ denotes the probability per unit time for a particular process to occur. 
This dynamics conserves the total number of $H$ (and $L$) particles. Under the
weight of the particles, a local hill on the surface gets pushed downward, 
at a higher rate by $H$ than by $L$. In one dimension, the surface consists of a 
chain with $N$ sites. The lattice bonds representing discrete surface elements, can 
have two possible orientations with slopes $\tau_{i+1/2} = \pm 1$, which are
called upslope and downslope bonds, respectively. In one dimension surface 
dynamics can be represented as
\begin{eqnarray}
W(/ H \backslash \rightarrow \backslash H /)&=& E+b \nonumber\\
W(\backslash H / \rightarrow / H \backslash)&=& E-b \nonumber\\
W(/ L \backslash \rightarrow \backslash L /)&=& E-b' \nonumber\\
W(\backslash L / \rightarrow / L \backslash)&=& E+b' 
\label{eq:model}
\end{eqnarray}
This dynamics conserves the overall slope. In two dimensions the surface consists 
of a square lattice and height of a site can change, provided all four 
neighboring sites are at the same height \cite{2dpassive,sup}.
We consider periodic boundary conditions with  no overall tilt of the surface.

\begin{figure}
\includegraphics[scale=1.0]{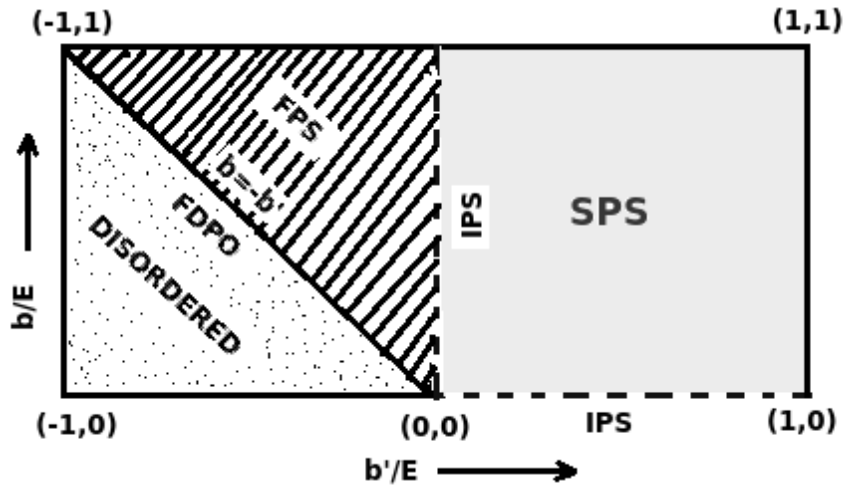}
\caption{Phase diagram in the $b-b'$ plane. For $b>0$ and $b'>0$, the system
shows SPS. The dotted horizontal and vertical lines are 
related to each other via interchange of the two particle species. 
On these lines the system is in IPS phase. The striped region ($-b<b'<0$)
represent FPS phase. $b=-b'$ line corresponds to FDPO phase. The dotted region
($b'<-b$) correspond to disordered phase.}
\label{fig:phase}
\end{figure}

Figure \ref{fig:phase} shows the phase diagram of the system in $bb'$ plane,
with $a$ taken to be positive. It follows from Eq. \ref{eq:model} that
interchanging $b$ and $b'$ is tantamount to interchanging the $H$ and $L$
species. Hence we consider positive $b$, while $b'$ can be positive or negative 
or zero.

{\it Strong Phase Separation (SPS)}: The right half of the phase diagram, $b' > 0$,
corresponds to the LR model \cite{lahiri97, lahiri}, as appropriate to sedimenting 
colloidal crystals \cite{lahiri97}. In this regime, the light particles tend to move 
the surface upward. In steady state, the upslope and downslope surface bonds 
phase separate to form a single macroscopic valley and hill, which hold 
all the $H$ and $L$ particles, respectively, in  separated clusters. Both 
particles and tilts show SPS; the approach to steady
state is extremely slow (involving times $T_{relax} \sim \exp(\alpha N) )$ 
owing to the formation of large metastable barriers. 

In the left half of the phase diagram in Fig. \ref{fig:phase} we have 
$b'<0$, which means
that the part of the surface occupied by $L$ particles shows a downward drift.
As discussed below, 
different phases are obtained depending on whether this velocity is larger than, 
smaller than, or the same as the velocity imparted by the $H$ particles. 

{\it Disordered Phase}: When the $L$ particles push the surface faster (dotted
region in  Fig. \ref{fig:phase}), neither of the
particle species nor the tilts shows long-range order. Interesting dynamical 
aspects of this wave-carrying disordered phase will be presented elsewhere 
\cite{prep}.

{\it Fluctuation-dominated Phase Ordering (FDPO)}: 
On the line $b=-b'$, both $L$ and $H$ particles push the 
landscape down at the same rate. This implies Kardar-Parisi-Zhang dynamics for the surface while
$L$-$H$ exchange rules (Eq. \ref{eq:model1}) imply that $H$-particles tend to 
collect in
local valleys. This reduces to the passive scalar problem studied earlier, in
which particles exhibit FDPO, characterized by singularities of the 2-point
correlations and giant fluctuations of the density \cite{fdpo,kapri}. 
Interestingly, this phase boundary can be identified exactly by looking for the 
onset of complex eigenvalues in a linear stability analysis of the coupled 
hydrodynamic equations for the landscape and the particles \cite{sup}.

{\it Fast-fall with Phase Separation (FPS)}: When $L$ particles push the surface
down at a slower rate than $H$'s, a new phase ensues (shown striped in 
Fig. \ref{fig:phase}). In steady state, $L$ and $H$ particles separate completely 
as in SPS. The surface underlying the $H$-cluster forms a macroscopic 
valley but unlike SPS, the phases are not pure, {\sl e.g.} the macroscopic 
majority-upslope region accommodates a finite fraction $(1-m)$ of downslope bonds. 
The majority-upslope
region in turn acts as a `tilt reservoir' which drives a finite tilt current through
the part of the surface which holds $L$ particles. By identifying an upslope 
(downslope) bond with a particle (hole) we identify the phase as the maximal
current phase in an open-chain asymmetric exclusion process \cite{bdry,sup}. 
Consequently,   
 near the edges of the $L$-domain the tilt density $\rho$ shows an algebraic 
$1/\sqrt{r}$ variation, while in the bulk $\rho \approx 1/2$ \cite{sup}.
Equating the tilt current $J=2bm(1-m)$ in the two arms holding the $H$ particles
with that in the maximal current phase $J'=2b'/4$ in the $L$-rich portion, we
deduce $4m(1-m)=b'/b$, a relation we have verified numerically.   
The presence of a finite tilt current through the system results in
a finite downward velocity of the surface and in steady state, the full surface
moves downward at finite speed, 
preserving the macroscopic valley and disordered tilt region, 
along with the pure domains of $H$ and $L$ particles (Fig. \ref{fig:scheme}a).

{\it Infinitesimal-fall with Phase Separation (IPS)}: For $b'=0$ 
(the vertical dashed line in Fig. \ref{fig:phase}) the local
fluctuations in the surface occupied by $L$ particles are of the symmetric
Edwards-Wilkinson type. In this phase, the $H$ and $L$ particles again
form pure domains. The surface beneath the $H$-cluster has the shape of a deep
valley consisting of pure domains of upslope and downslope bonds. By contrast, 
the surface
occupied by the $L$-cluster in this case behaves like an open-chain symmetric
exclusion process connected to the two reservoirs of upslope and downslope tilts
at the two ends. Thus the tilt density varies linearly in this region with a
gradient $\sim 1/N$ \cite{derrida_sep,sup}, leading to a tilt current and an 
infinitesimal downward velocity $\sim 1/N$ of the entire landscape. We 
schematically show a typical configuration in Fig. \ref{fig:scheme}a.

The general properties of the phases discussed above remain valid even in two 
dimensions, where $H$ and $L$ particles form compact clusters. The shape of these 
two dimensional clusters depends on the topography of the surface heights. As in
one dimension, we find a deep valley that holds the $H$-cluster. Measured from the
bottom-most point, the height increases linearly in both 
$x$ and $y$ directions and it is easy to see that in this case the equal 
height contours have the shape of a diamond. In Fig. \ref{fig:scheme}b and 
\ref{fig:scheme}c, we show some representative configurations.

\begin{figure}
\includegraphics[scale=0.5,angle=0]{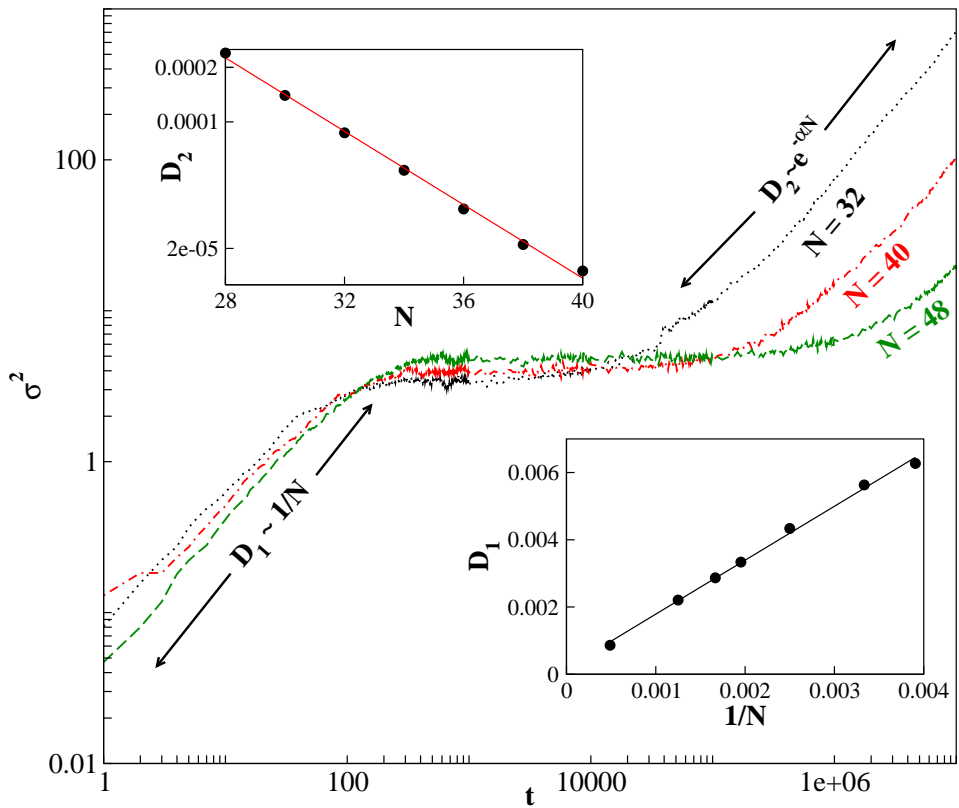}
\caption{Occurrence of three regimes in the steady state dynamics of the deepest 
point of the valley. Main plot: Mean squared displacement of the deepest point of the
valley as a function of time. The displacement shows an initial diffusive growth, 
followed by a plateau, and finally another diffusive
regime at large time.  Bottom Inset: Small time diffusivity $D_1 \sim 1/N$.
Top Inset: For large time, the diffusivity $D_2 \sim e^{-\alpha N} $, with
$\alpha \simeq 0.26$. These data correspond to $b'=0$, $b=E$, $a=D$ 
and have been averaged over $5000$ steady state configurations.} 
\label{fig:msd}
\end{figure}

The way in which the landscape is organized in the IPS and FPS phases has a
profound influence on dynamical properties. For instance, although the centre of
mass of the $H$ cluster remains stationary for a long time, 
the landscape immediately below it
undulates in time, leading to three distinct temporal regimes in steady state.
These are captured by monitoring the mean-squared displacement $\sigma^2$ of the 
deepest point of the valley. Figure \ref{fig:msd} presents the data for $b'=0$. At small
time $t \ll N^2$, we find $\sigma^2$ grows diffusively with a diffusion constant 
$D_1 \sim 1/N$. But after times of order $N^2$, a plateau for $\sigma^2$ is reached 
at a value $\sim N$. From a simple consideration of the total (gravitational) energy
of the $H$ particles, it is easy to show that when the deepest point coincides with 
the centre of mass of the $H$ cluster, the energy is minimum. Any displacement from 
this position gives rise to a restoring force that scales linearly with the 
displacement. The motion of the deepest point is thus described by an 
Ornstein-Uhlenbeck process \cite{orn}; consequently, the deepest point diffuses 
within a region of size $\sqrt{N}$  around the $H$ cluster centre of mass \cite{sup}. 
Finally, at very large $t$, the $H$ cluster itself moves diffusively around the 
system and the valley moves along with it (see \cite{sup}). The 
mean-squared displacement of the deepest point in this regime has a 
diffusion coefficient $D_2 \sim e^{- \alpha N}$.

\begin{figure}
\includegraphics[scale=0.5,angle=0]{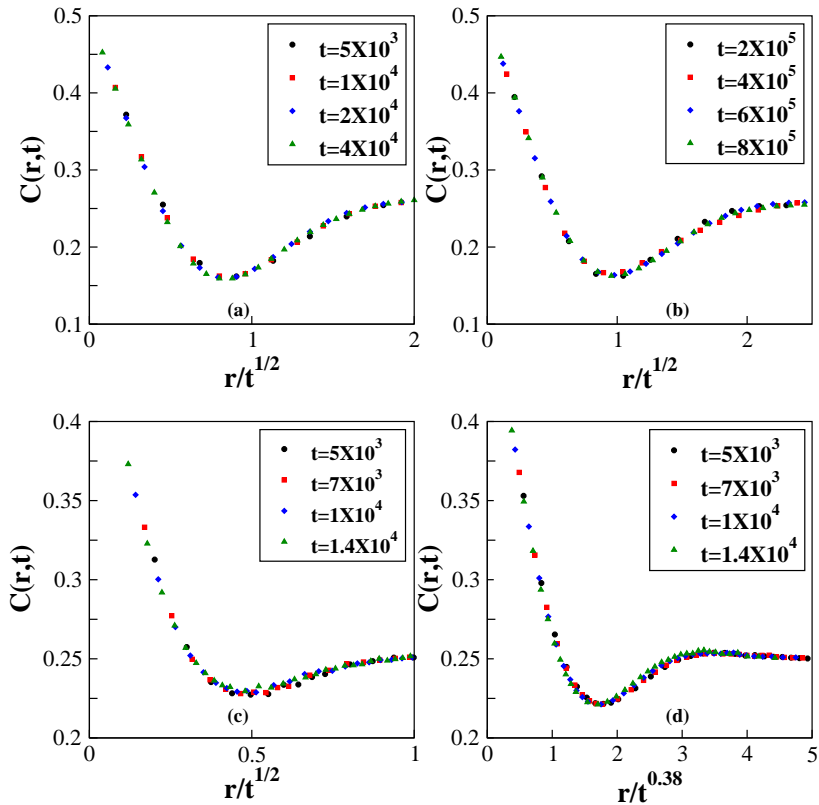}
\caption{ Scaling of particle density correlation in the coarsening phase. 
in both one and two dimensions. The equal time density correlation for the 
particles $C(r,t)$ shows a collapse when $r$ is scaled by 
${\cal L}(t) \sim t^{1/z}$. (a) and (c) show scaled data for $b=0.5, b'=0$ in one and
two dimensions, respectively. Here we find $z \simeq 2$. (b) and (d) show 
similar plots for $b=0.3, b'=-0.2$. Here, in one dimension, $z \simeq 2$ and in two
dimensions $z \simeq 2.6$. We have used $N=1024$(a), $N=16384$(b) and $N=256\times256$((c) and (d)) here.}
\label{fig:rel}
\end{figure}

Another important aspect of the dynamics concerns the relaxation to steady state
starting from an initially disordered state. Interestingly, IPS and FPS
phases show an enormous reduction in this relaxation time, as compared to 
earlier known examples of SPS as in the LR and ABC models \cite{lahiri,abc}. 
For $b'>0$ (LR model), 
the landscape occupied by an $L$-cluster tends to move upwards and forms a hill, 
while an $H$-cluster pushes the landscape down and forms a valley. 
For two adjacent valleys with $H$-clusters to merge, the time to dissolve the
intermediate hill containing the $L$-cluster grows exponentially with the size of the
$L$-cluster, and hence the final SPS state is reached over a time-scale
$e^{\gamma N}$. By contrast, in the FPS and IPS ($b' \le 0$) phases, the landscape 
is organized differently, and this leads to fast relaxation, with times growing as 
$N^z$. This is because the part of the landscape beneath the intervening $L$ 
cluster either shows symmetric fluctuations (for $b'=0$) or gets pushed downward 
(for $b'<0$). Figure \ref{fig:rel} shows the scaling
collapse of the equal time density correlation function for $H$ particles when
separations are scaled by the coarsening length scale ${\cal L} (t)$, which is 
found to grow as $t^{1/z}$. In IPS phase ($b'=0$) we find $z=2$ in both one and two
dimensions. The FPS phase ($-b<b'<0$) shows $z \simeq 2$ in one dimension for 
very large 
$N$ and $t$ while for smaller values of these variables, our data show finite size 
effects (see \cite{sup} for details). In two dimensions for FPS, we measured 
$z \simeq 2.6$, for the largest values of $N$ and $t$ we could access.
Our plots in Fig. \ref{fig:rel} constitute the first observation of 
algebraic coarsening for completely phase separated systems, and stand in strong 
contrast to the ultraslow logarithmic coarsening observed in the LR and ABC models 
\cite{lahiri,abc}. Underlying the speed-up of the coarsening process is a 
simple mechanism, namely the formation of disordered segments of the landscape 
between ordered clusters. These segments generate fluctuations which allow 
mergers of ordered regions to occur on a rapid time scale. 

To summarize, we have considered coupled driven systems consisting of two
species of particles being advected by an energy landscape whose dynamics is in
turn influenced by the particles. The differential interaction between the
landscape and the two particle species gives rise to different phases in the
system as the interaction parameters are varied. We have demonstrated the 
occurrence of new phases with fast dynamics, where both the particles and the 
landscape show long range
order and the composite system is in a nonequilibrium current-carrying state.

We conclude with a discussion of the implications of our work for modeling in 
biophysical contexts. It is known that on the cell membrane, various
proteins and lipids like integrins, T-cell receptors are present in the form 
of nano-clusters \cite{zanten,yu}. This clustering is shown to be induced by 
the cortical actin cytoskeleton \cite{madan08,madan12}, and within a recent 
theoretical model of the process, an FDPO state has been observed \cite{madan16}.
There is now experimental evidence that the  actin cytoskeleton also gets 
reorganized by these membrane components \cite{varma,yu}.  This 
raises the interesting possibility of new phases arising if the treatment of 
\cite{madan16} is extended to account for the back action of membrane components on
the cytoskeleton.

We acknowledge useful discussions with T. Sadhu, M. Rao, and A. Das. 
The computational facility used in this work was
provided through Thematic Unit of Excellence on Computational Materials Science,
funded by Nanomission, Department of Science and Technology, India.


\appendix

\setcounter{equation}{0}

\setcounter{figure}{0}

\section{Simulation details in two dimensions and finite size effects}

\setcounter{equation}{0}

\setcounter{figure}{0}
The surface in two dimensions is simulated through a discrete solid-on-solid 
algorithm, where the height difference between the nearest neigbours on a 
square lattice is maintained at $\pm 1$. Let $h(i,j)$ denote the height of 
the site $(i,j)$ on the lattice. This site is said to be on a local hill, if
all its four neigbours with coordinates $(i\pm 1, j)$ and $(i,j \pm 1)$ have 
height $h(i,j)-1$.
Similarly, the site $(i,j)$ is said to be in a valley when the neighbors
have height $h(i,j)+1$. The two dimensional surface evolves in time
by switching between the hills and the valleys. In our model, a site is selected
at random. If it is on a hill, then it can flip to a valley when its height gets 
reduced by two units. This flipping rate is $(E+b)$ when the site is occupied by 
an $H$ particle and $(E-b')$ when it is occupied by an $L$ particle. Similarly,
if the chosen site happens to be in a valley, then its height can increase by two
units and it becomes a hill. The switching rate in this case is $(E-b)$ when an $H$
particle is present, and $(E+b')$ when an $L$ particle is present. For updating the 
particles, we select a bond (horizontal or vertical) of the square lattice. If
the two sites adjacent to the bond are occupied by two different species of 
particles, then their positions are exchanged with rate $(D+a)$ if after the 
exchange, height of the $H$ particle decreases. The reverse exchange occurs with
rate $(D-a)$.

\begin{figure}[h]
\includegraphics[scale=0.5,angle=0]{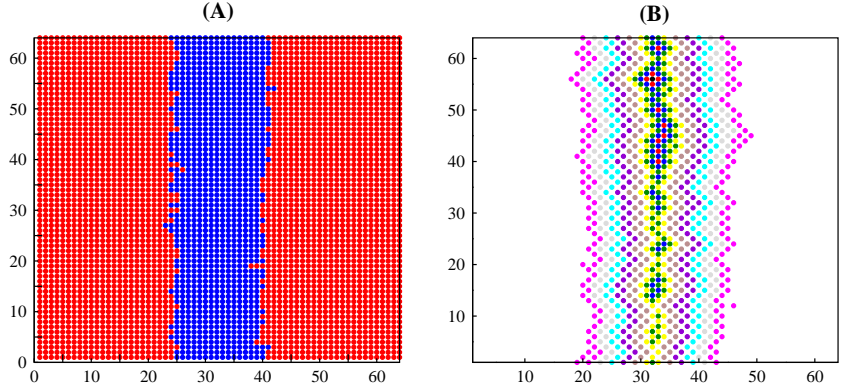}
\caption{Representative plots for wedge type configurations 
in the IPS phase in two dimensions 
on a $64\times64$ square lattice. In (A) $H$($L$) clusters are shown by 
red(blue) color, while (B) shows equal height contour plots.}
\label{fig:scheme2}
\end{figure}

In our simulation, we start with a flat initial
configuration of the surface and random configuration for the particles and 
evolve the system in time, following the algorithm described above. In steady 
state the surface develops a single valley where the equal height contours 
resemble diamonds, shown in Fig. 1c in the main paper. However, in certain cases, 
we also observe another type of state, where a completely different height
topography emerges. Instead of a single minimum seen for diamond shaped 
contours, we find a line of minima and height increases along the direction
perpendicular to this line. In other words, the surface has the shape of 
a trench. The equal height contours and the perimeter of the $H$-cluster
for such a configuration is shown in Fig. \ref{fig:scheme2}. We have
verified that such wedge-like configurations result from finite size effects in
our system with periodic boundary conditions 
and for larger systems only diamond-shaped contours are found.  

\newpage
  

\section{Linear stability analysis of the coupled hydrodynamic equations}

\setcounter{equation}{0}

\setcounter{figure}{0}
From the dynamical rules shown in Eqs. 1 and 2 in main text, the particle
current and the tilt current can be written as
\begin{align}
J_{\rho} &= 2a\rho(x,t)(1-\rho(x,t))(1-2m(x,t)) \nonumber \\
J_m &= m(x,t)(1-m(x,t))[2\rho(x,t)(b+b^\prime)-2b^\prime]
\label{sys}
\end{align}
$\rho(x,t)$ and $m(x,t)$ denote the coarse-grained densities of the $H$ particles and the 
upslope bonds respectively. We write each density as sum of the average
value and a small perturbation:  $\rho(x,t) = \rho_0 + \delta \rho(x,t)$ and 
$m(x,t) = m_0 + \delta m(x,t)$ and expand currents in Eq. \ref{sys} upto linear
order in $\delta \rho$ and $\delta m$. This gives us the continuity equation
\begin{equation}
\frac {\partial \vec{u}}{\partial t} + A \frac{\partial\vec{u}}{\partial x} =0
\end{equation}
where we have used the compact notation $\vec{u}$, denoting a column vector with 
components $\rho$ and $m$. The matrix $A$ is defined as 

\[
A=
  \begin{bmatrix}
   \frac{\partial J_{\rho}}{\partial \rho} ~~~ \frac{\partial J_{\rho}}{\partial m} \\
    \frac{\partial J_{m}}{\partial \rho} ~~~ \frac{\partial J_{m}}{\partial m}
  \end{bmatrix} = \begin{bmatrix}
   \hspace{-85pt}2a(1-2\rho_0)(1-2m_0) ~~~~~~-4a\rho_0(1-\rho_0) \\
   2m_0(1-m_0)(b+b^\prime) ~~~~~~ 2\rho_0(1-2m_0)(b+b^\prime) -2b^\prime(1-2m_0).
  \end{bmatrix}
\] 
For an untilted surface, $m_0=1/2$, and the eigenvalues are
\begin{equation}
\lambda= \pm \sqrt{-2a\rho_0(1-\rho_0)(b+b^\prime)}
\end{equation}
For $b>-b^\prime$, one has imaginary eigenvalues, indicating linear instability. 
In this case, the system shows macroscopically ordered phases. For $b <-b^\prime$, 
the eigenvalues are real and the system is disordered. 
The $b= -b^\prime$ straight line separates the ordered and the disordered phase 
and on this line, the system exhibit FDPO.
\section{Static Correlations in Steady State}

\setcounter{equation}{0}

\setcounter{figure}{0}

\begin{figure}[h]
\includegraphics[scale=0.5,angle=0]{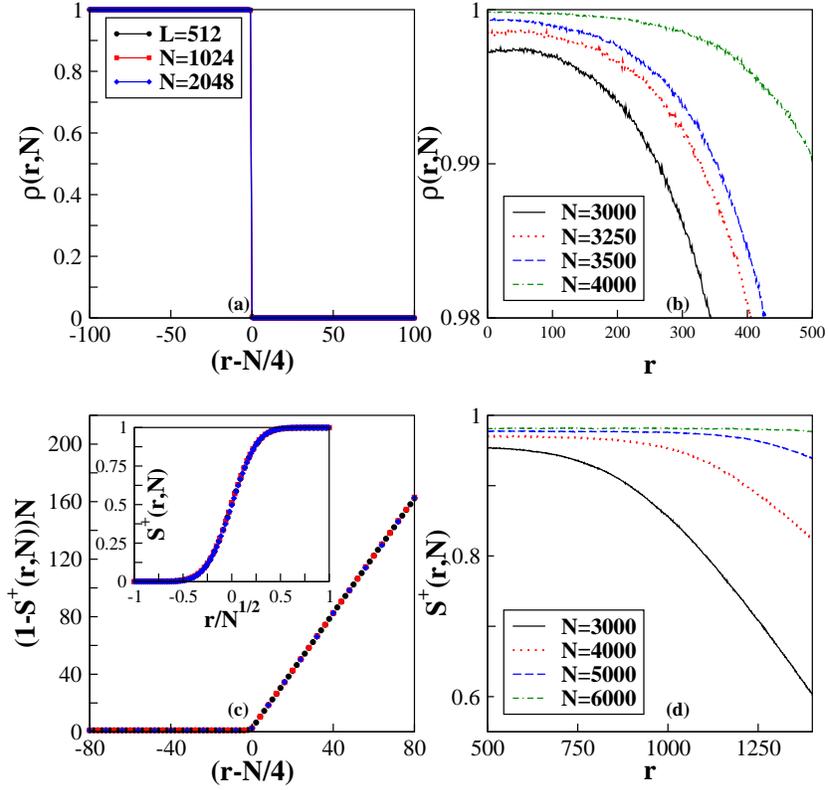}
\caption{Static correlation functions in IPS phase. (a): Density of $H$ particles 
$\rho(r,N)$ at $q=0$. The density changes sharply from $1$ to $0$ indicating
pure phases of $H$ and $L$ particles. (b): For $q=(D-a)/(D+a)=0.99$ (
see Eq. 1 in the main text), $H$ particle density
approaches $1$ as $N$ is increased. This confirms the existence of a pure phase 
in the thermodynamic limit for all $q<1$. (c) Main plot: For $r \gg \sqrt{N}$ a 
scaling collapse is obtained for different system sizes when $[1-S^+(r,N)]N$ is 
plotted against  $(r-N/4)$. Inset shows that as $r$ changes sign, $S^+(r,N)$ 
shows a transition from $0$ to $1$ across the domain boundary, which
is at the deepest point of the valley. Due to diffusive motion of the domain
boundary, $S^+(r,N)$ is a scaling function of $r/\sqrt{N}$ in this region. 
In main plot and inset we show data for $N=512$ (black circles), $1024$ (red
squares) and $2048$ (blue diamonds). (d): For $q=0.99$ and $0\ll r\ll N/4$ 
 $S^+(r,N) \to 1$ as $N$ is increased. 
These data have been averaged over at least $3 \times 10^5$ steady state 
configurations.}
\label{fig:static}
\end{figure}

\begin{figure}[h]
\includegraphics[scale=0.5,angle=0]{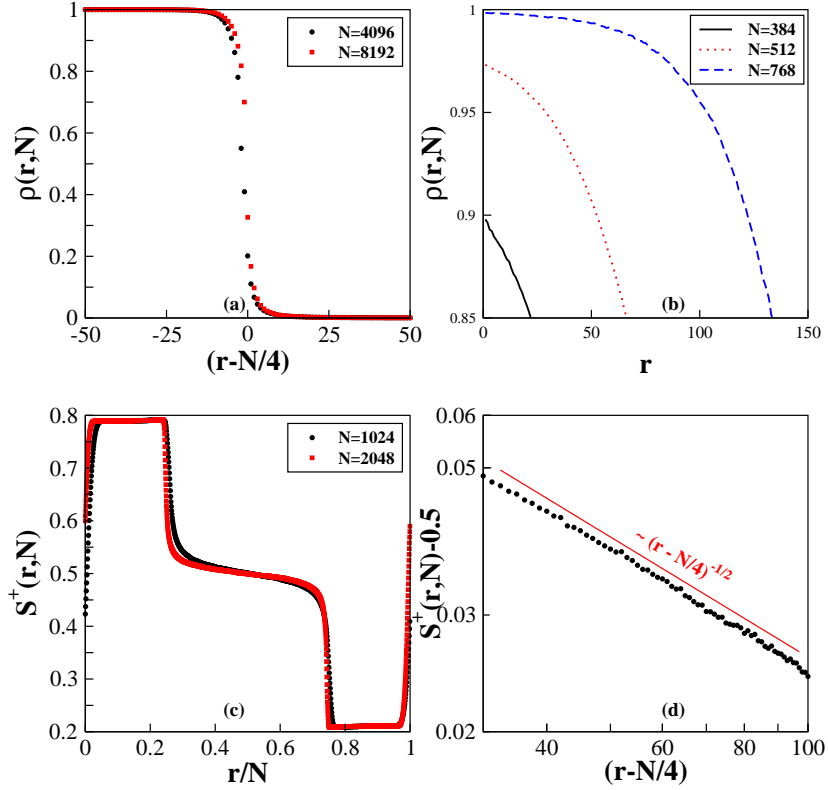}
\caption{Static correlation functions in FPS phase with $b=0.3$ and $b'=-0.2$.
(a): Density of $H$ particles $\rho(r,N)$. The density changes sharply from $1$ 
to $0$ indicating pure phases of $H$ and $L$ particles. (b): For $q=(D-a)/(D+a)=
0.92$ (see Eq. 1 in main text), $\rho(r,N)$ approaches $1$ as $N$ is increased.
This demonstrates existence of pure $L$ and $H$ domains in the thermodynamic 
limit, as long as $q <1$, {\sl i.e.} the 
$H$ particles preferentially slide downward. (c): $S^+(r,N)$ for small $r$
remains constant at a value which is less than unity. This indicates that the
upslope and downslope bonds underlying $H$-cluster undergo phase separation but
no pure phases are formed. The variation of $S^+(r,N)$ in the region occupied 
by $L$ cluster shows an algebraic decay to a disordered phase, as expected for
maximal current phase of an open-chain ASEP. (d): The algebraic decay of $S^+(r,N)$
occurs with an exponent $1/2$, consistent with earlier known results for maximal
current phase. These data have been averaged over at least $10^6$ steady state 
configurations.}
\label{fig:static_striped}
\end{figure}

In Figs \ref{fig:static} and \ref{fig:static_striped}, we present our results
on the static correlation functions for the particles and for the surface bonds
in one dimension. 
Figure \ref{fig:static} corresponds to IPS phase and 
Fig. \ref{fig:static_striped} correspond to FPS phase. Here we have considered
equal number of $H$ and $L$ particles and all distances have been measured from
the centre of mass of the $H$-cluster. $\rho(r,N)$ and $S^+(r,N)$ denote, 
respectively, the density of $H$ particles and the probability to find an upslope 
bond at a distance $r$ from the centre of mass.

\newpage

\newpage
\section{Mechanisms responsible for two distinct time scales in valley dynamics}

\setcounter{equation}{0}

\setcounter{figure}{0}

\begin{figure}[h]
\includegraphics[scale=0.4]{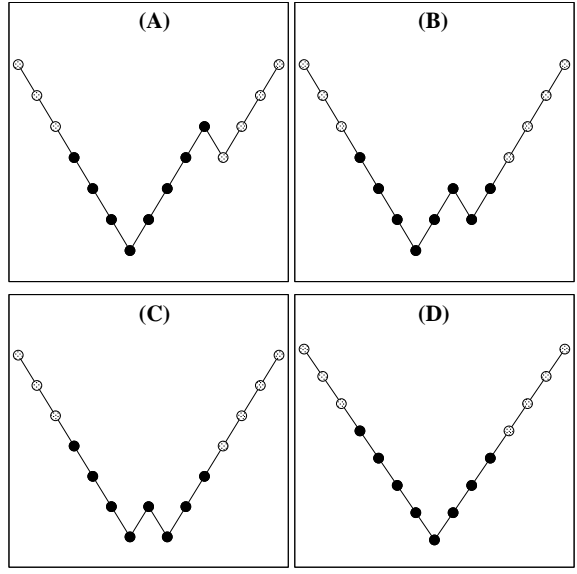}
\caption{Motion of the deepest point over $\sim N^2$ time-scale.
(A) An occupied local hill forms at the right edge of the particle
cluster. (B) The occupied local hill, being unstable, zips through the pure
domain of upslope bonds. (C) The local hill reaches the bottom of the valley.
(D) The deepest point shifts one site to the right.}
\label{fig:smallt}
\end{figure}

In this section we explain the mechanisms for the short and large time valley dynamics. 
The deepest point of the valley lies at the boundary between two pure domains
of upslope and downslope surface bonds and these pure domains extend upto the
edges of the $H$ particle cluster. Beyond those edges,  
$S^+(r,N)$ has a linear gradient of order $1/N$ and it is easy to show by a simple 
mean-field calculation that the average distance of the nearest downslope (upslope) 
bond from the right (left) edge of the particle cluster  $\sim \sqrt{N}$.
Therefore this  downslope (upslope) bond can diffuse through this distance
over a time-scale $\sim N$, and 
form a local hill, as shown in Fig. \ref{fig:smallt}A and then the downslope
(upslope) bond  moves ballistically down the valley, over a time-scale $\sim
N$, and reaches the deepest point causing a unit displacement of the deepest
 point towards right (left). Thus 
the process illustrated in Fig. \ref{fig:smallt} takes place over a time-scale
$\sim N$  which explains the $D_1 \sim 1/N$ behavior for short times. The 
mechanism also shows that
 the particles do not undergo any displacement in this case. The deepest
point of the valley slides back and forth beneath the $H$ particle cluster.
This motion, however has an energy cost, and as discussed in the main text, for
a distance $\Delta$ between the deepest point and the centre of mass of the
$H$ cluster, the energy cost $\sim \Delta ^2$. Thus the motion of the valley
can be described by an Ornstein-Uhlenbeck process. This prevents the valley from 
taking a large excursion in either direction and as a result the mean-squared 
displacement reaches a plateau.

\begin{figure}[h]
\includegraphics[scale=0.4,angle=0]{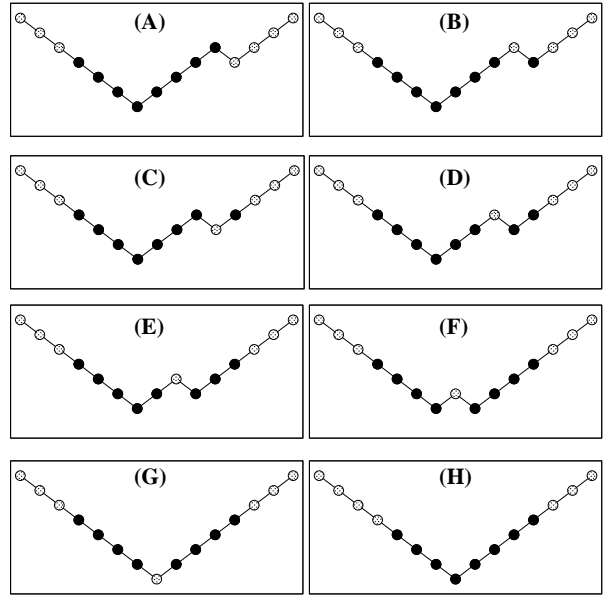}
\caption{Motion of the deepest point over $\sim e^N$ time-scale.
(A) Formation of a local hill at the right edge of the $H$ particle cluster.
(B) The rightmost $H$ particle slides down the hill and gets detached from the
cluster. (C) The resulting hill with $L$ particle flips and another local hill
with $H$ particle is
formed at the adjacent site. (D) Another $H$ particle slides down the hill and
detaches from the $H$ cluster. (E) The $L$ particle propagates down the valley. 
(F) At the
bottom of the valley a local hill with $L$ particle is formed. (G) This local hill
flips and the deepest point of the valley shifts one site to the right.
(H) The $H$ particles to the left of the deepest point slide down, one after
another, moving the center of mass of the particle cluster to one site to the
right.}
\label{fig:larget}
\end{figure}

\begin{figure}[h]
\includegraphics[scale=0.4]{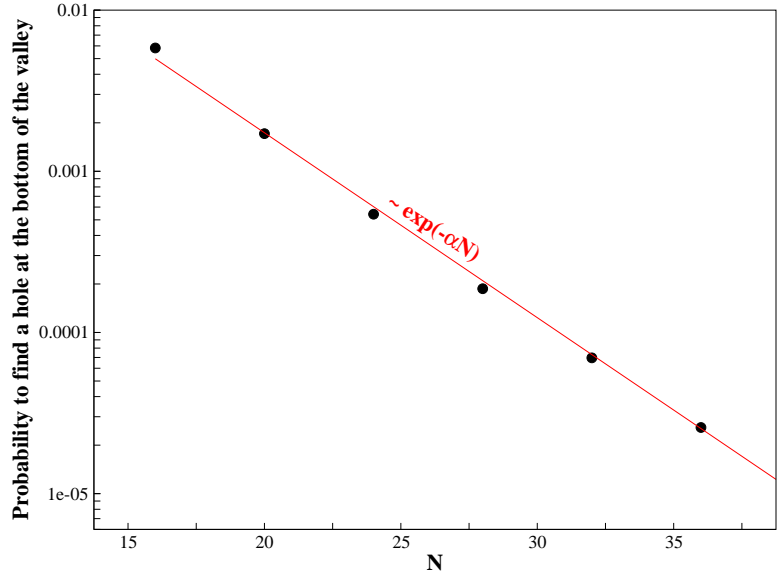}
\caption{The probability to find an $L$ particle at the deepest point of the surface 
falls off as $e^{-\alpha N}$ with system size where $\alpha=0.261$. The data
have been averaged over  $10^6$ histories.}
\label{hole}
\end{figure}

In the limit of very long time, the $H$ particle cluster will start
moving around the system diffusively and the valley will naturally move along
with it. This mechanism has been illustrated in Fig. \ref{fig:larget} which
shows movement of an $L$ particle through the $H$ particle cluster. 
When the $L$ particle reaches
the bottom of the valley, the deepest point undergoes a displacement, along
with the center of mass of the $H$ particle cluster. Note that the time-scale for
this process is rather large because the probability that the $L$ particle, 
starting from the $HL$ domain boundary, reaches the bottom of the
valley, is very low and decays exponentially with the domain size (see our
data in Fig. \ref{hole}). As a result, the diffusivity of the valley in this regime
$\sim e^{-\alpha N}$. Existence of an exponentially large time-scale breaks the
translational invariance of the system but the relaxation time-scale is still
algebraic, as demonstrated in our main text.     

\section{Dynamical exponent for the FPS phase}

\setcounter{equation}{0}

\setcounter{figure}{0}
Two point density correlation function for the $H$ particles is defined as 
\begin{equation}
C(r,t)=<\rho_i(t)\rho_{i+r}(t)> - \rho_0^2
\end{equation}
where, $\rho_i(t)$ is 1(0) if the $i^{th}$ site is occupied by an $H$($L$) particle at time $t$ during the coarsening phase. The angular brackets denote average over initial
conditions. $\rho_0$ denotes the global density of $H$ particles in the system.
Define $r_0$ as $C(r_0,t)=0$ and the dynamical exponent $z$ can be estimated
from the variation of $r_0$ with $t$, using the scaling relation $r_0 \sim t^{1/z}$. 
The plots of $r_0$ vs $t$ for FPS phase in one and two dimensions  
are presented in Figs \ref{fig:fps1d} and \ref{fig:fps2d}, respectively. 
\newpage
\begin{figure}[h]
\includegraphics[scale=0.5]{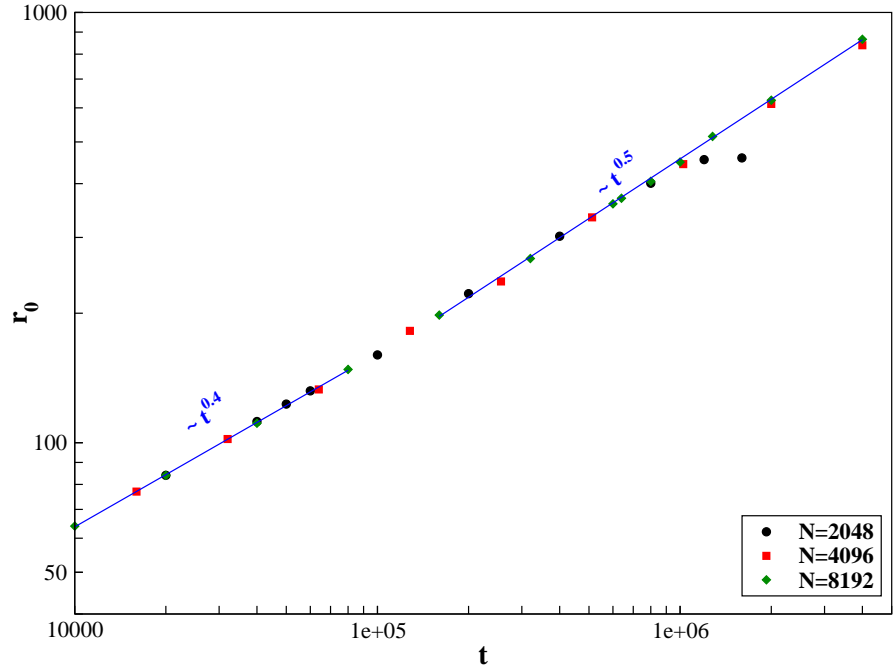}
\caption{Plot of $r_0$ vs time during the coarsening phase for $b=0.3$,
$b^\prime=-0.2$ in one dimension. For large N and t, we obtain
$r_0 \sim t^{0.5}$. For $t \lesssim 10^5$, we observe a crossover
region where $r_0 \sim t^{0.4}$. We have used $\rho_0=1/2$ here. 
\label{fig:fps1d}
 }
\newpage
\end{figure} 
\begin{figure}[h]
\includegraphics[scale=0.5]{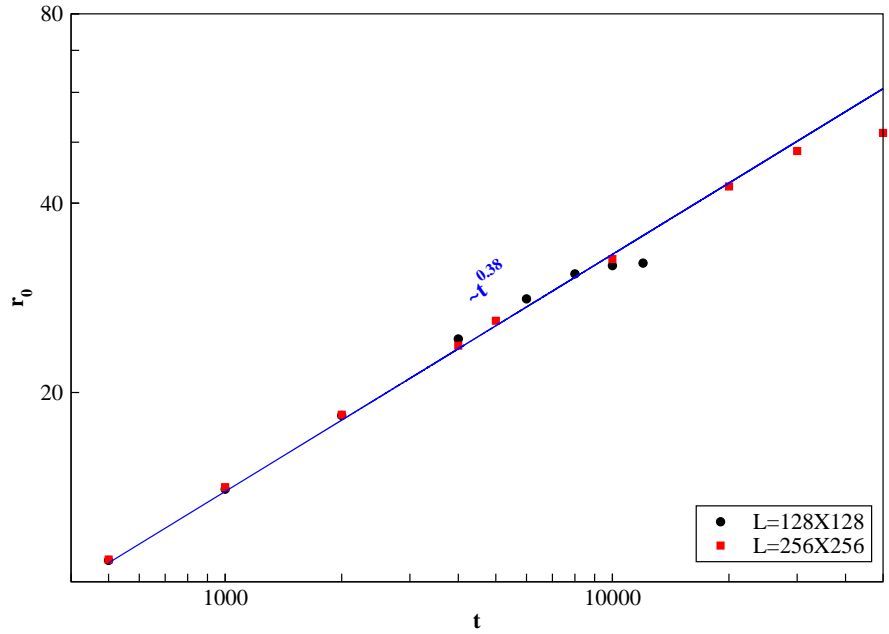}
\caption{Plot of $r_0$ vs t during the coarsening phase for $b=0.3$, 
$b^\prime=-0.2$ in two dimensions.
We obtain $r_0 \sim t^{0.38}$. We have used $\rho_0=1/8$ here. 
\label{fig:fps2d}
 }
\end{figure}
\end{document}